\documentclass[sigconf]{acmart}

\AtBeginDocument{%
  \providecommand\BibTeX{{%
    \normalfont B\kern-0.5em{\scshape i\kern-0.25em b}\kern-0.8em\TeX}}}

\usepackage{url}

\setcopyright{acmcopyright}

\copyrightyear{2023}
\acmYear{2023}
\setcopyright{rightsretained}
\acmConference[WWW '23]{Proceedings of the ACM Web Conference 2023}{April 30-May 4, 2023}{Austin, TX, USA}
\acmBooktitle{Proceedings of the ACM Web Conference 2023 (WWW '23), April
30-May 4, 2023, Austin, TX, USA}
\acmDOI{10.1145/3543507.3587432}
\acmISBN{978-1-4503-9416-1/23/04}

\begin{document}
\newcommand{\rza}[1]{{\color{red} #1}}

\title{Tangible Web: An Interactive Immersion Virtual Reality Creativity System that Travels Across Reality}


\author{Simin Yang}
\authornotemark[1]
\affiliation{%
  \institution{Hong Kong University of Science and Technology}
  \city{Hong Kong SAR}
  \country{China}}
\email{syangcj@connect.ust.hk}

\author{Ze Gao}
\authornote{Both authors contributed equally to this research.}
\affiliation{%
  \institution{Hong Kong University of Science and Technology and Hong Kong University of Science and Technology (Guangzhou)}
  \city{Hong Kong SAR and Guangzhou}
  \country{China}}
\email{zgaoap@connect.ust.hk}

\author{Reza Hadi Mogavi}
\affiliation{%
  \institution{Hong Kong University of Science and Technology and University of Waterloo}
  \city{Hong Kong SAR and Waterloo}
  \country{China and Canada}}
\email{rhadimogavi@acm.org}

\author{Pan Hui}
\affiliation{%
  \institution{Hong Kong University of Science and Technology (Guangzhou)}
  \city{Guangzhou}
  \country{China}}
\email{panhui@ust.hk}

\author{Tristan Braud}
\authornote{Tristan Braud is the corresponding author.}

\affiliation{%
  \institution{Hong Kong University of Science and Technology}
  \city{Hong Kong SAR}
  \country{China}}
\email{braudt@ust.hk}

\renewcommand{\shortauthors}{Yang and Gao, et al.}

\begin{abstract}


With the advancement of virtual reality (VR) technology, virtual displays have become integral to how museums, galleries, and other tourist destinations present their collections to the public. However, the current lack of immersion in virtual reality displays limits the user's ability to experience and appreciate its aesthetics. This paper presents a case study of a creative approach taken by a tourist attraction venue in developing a physical network system that allows visitors to enhance VR's aesthetic aspects based on environmental parameters gathered by external sensors. Our system was collaboratively developed through interviews and sessions with twelve stakeholder groups interested in art and exhibitions. This paper demonstrates how our technological advancements in interaction, immersion and visual attractiveness surpass those of earlier virtual display generations. Through multimodal interaction, we aim to encourage innovation on the Web and create more visually appealing and engaging virtual displays. It is hoped that the greater online art community will gain fresh insight into how people interact with virtual worlds as a result of this work.

\end{abstract}

\begin{teaserfigure}
  \includegraphics[width=\textwidth]{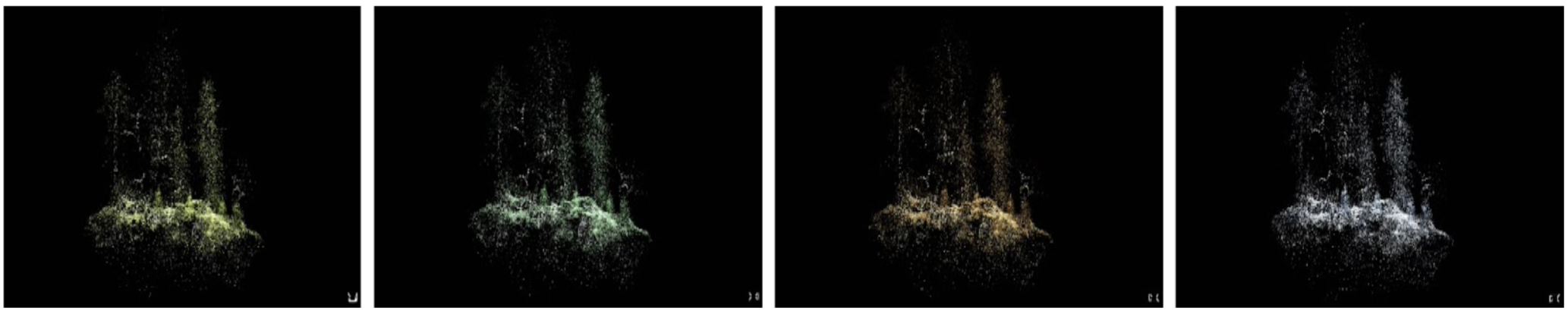}
  \caption{Four Seasons and their corresponding temperatures measured through sensors attached to VR devices (10℃ to 30℃). From left to right: spring (Light green), summer (dark green), autumn (dark yellow), and winter (white).}
  \Description[Four Seasons and their corresponding temperatures measured through sensors attached to VR devices (10℃ to 30℃).]{From left to right: spring (Light green), summer (dark green), autumn (dark yellow), and winter (white).}
  \label{fig:teaser}
\end{teaserfigure}

\maketitle

\section{Introduction}

Since its creation in the early 1990s, the Web has  continually integrated richer media. Early webpages were primarily text- and image-based. Videos then became more pervasive thanks to improvements in network bandwidth. More recently, WebGL \cite{WebGL} enabled interactive 3D experiences over the web, quickly followed by WebXR for immersive experiences in Virtual and Augmented Reality (VR/AR). However, despite such technological advancements, immersive Web experiences tend to rely primarily on their novelty to attract users and do not seem to lead to prolonged interest.

WebXR \cite{webXR} enables users to rapidly try new XR experiences without installing an application on their devices by running on web browsers. Developing a single interoperable standard allows developers to focus on the application's content without considering the specifics of multiple platforms. However, such flexibility comes at the cost of significantly decreased performance compared to native XR applications. As such, web XR applications suffer from less attractive graphics than their native counterparts, leading to lower immersion. To compensate for such technological issues, we propose interfacing WebXR applications with other sensors to communicate a broader range of data to the application. Such data may include external environment info such as temperature and humidity, advanced user motion sensing, and interaction with physical objects and tangible interfaces. By integrating additional data on the users and their environments, WebXR can further blur the line between physical and digital, reinforcing immersion and offering a better user experience.

This paper presents a new artwork contribution based on a creative physical interaction system, "TangibleWeb". This physical interaction system can transfer real-world data to a VR world, empowering web VR with new capabilities and allowing users to interact with each other outside of the controller. 
"Tangible Web" is a virtual reality interactive device, which can be connected to various sensors (such as temperature, humidity, flame, etc.), and affect the content presentation in VR devices through sensor data. This way, the interaction between users and virtual artworks will diversify. For example, when users gradually warm the sensor with their hands, the scene in the VR device will also change with the temperature.

It will be realized by this method of traversing virtual and reality in a computer-programmed way, i.e., firstly, by sensing the change of outside temperature through temperature sensors (for example, by displaying matches prepared on site), secondly, receiving multi-sensor data through Arduino and send the data to the computer through the serial port, and finally, the virtual scene in the device that receives the command gradually changes from a silvery winter into a lush summer scene. 

This project aims to explore the physical and virtual connection through interactive technology. 
The virtual reality interactive installation can be seen as a bridge between virtual and reality through the system of the audience that can influence the virtual environment. In today's vigorous development of virtual reality technology, many studies are increasingly focused on the physical experience in a virtual environment. For example, according to Suh and Prophet's research \cite{suh2018state} on immersive VR at present, they put forward that the main areas of concern are Stimuli (i.e., sense and perception), Organism (i.e., cognition and emotion), Response (i.e., positive and negative results) and Individual differences in VR use (i.e., seeking personal cognition). But few studies and creations present and discuss the impact of reality on virtual environments~\cite{bhagavathula2018reality}. It is likely due to the limitations of head-mounted devices and controllers that limit diversified interaction. 


We will use this installation to provoke a discussion about the excessive invasion of media into urban and human life today \cite{kelly2017inevitable}. The public is thinking about virtual reality technology. It will be presented as an interactive installation, creating a new experience for the viewer to travel through reality and the virtual. Ideally, this project will give artists, researchers, and media researchers something to think about in the longer term, given the prevalence of metaverse technologies. For example, whether virtual reality should be more like the real world or compensate for what cannot be done in the real world.

Our contribution is:
\begin{itemize}
    \item We proposed a physical interaction system by transferring data created in the actual world (such as temperature, humidity, pressure, etc.) to the virtual world. We presented a physical interaction system in which the virtual world's environment is altered using data from the real world. Realize the relationship between virtual and actual worlds.

    \item More individuals may utilize this physical system to create creative experiences on the web by expanding the interface's capability through the data interface, which is not just restricted to the construction of button-based interaction.
\end{itemize}

\section{Formative Analysis and Design Decisions}
Following our university's Institutional Review Board (IRB) 's approval, we conducted semi-open-ended Zoom interviews with twelve participants: seven regular tourists, three artists, and two attraction managers from diverse nations and locations. The participants included three women and four men between the ages of 25 and 35. After analyzing the interviews, we reached the following conclusions:

\begin{itemize}
\item Several visitors mentioned that COVID-19 had prevented them from traveling for three years. They acknowledged that VR tours were an alternative but preferred to visit real places.
\item Based on the attraction administration's statements, considerable funds were allocated towards the 3D reconstruction and establishment of virtual reality (VR) tours during the outbreak. They wanted to ensure this investment does not go to waste.
\item Two female participants expressed a keen interest in interacting with and viewing real-world VR scenes that simulate various seasons and weather conditions. 
\item Consistent with the previous point, an artist recommended that an interactive experience would profoundly impact viewers.
\end{itemize}



Based on the above findings, we established the motivation for this study in the post-pandemic age. The increasing opening of actual tour attractions and the ongoing use of virtual settings inspired us to examine the link between the real and the virtual, including whether the virtual can replace the real or only enhance it.

The study explores design options that aim to enhance the user experience of VR tours, emphasizing participation, artistic merit, and genuine emotion.

\begin{itemize}
\item {\verb|Participation|}: User participation is crucial in improving user experience and encouraging repeat visits. By incorporating interactive elements, users can engage with the website on a deeper level, leading to a more memorable and enjoyable experience.
\item {\verb|The virtual scene's artistic merit|}: While virtual scenes do not need to replicate real-world scenes exactly, artistic processing can enhance the visual impact of the scene. After careful consideration, we utilized an algorithm to create a point cloud scene that met our artistic standards and captured the essence of the real-world scene.
\item {\verb|Genuine emotion|}: To create an authentic experience, we first carefully selected temperature and humidity sensors. Users can adjust the virtual scene by manipulating a water glass and a lighter image sensor, allowing them to simulate environmental changes and evoke genuine emotions.
\end{itemize}

\begin{figure*}[h]
    \centering
    \includegraphics[width=\textwidth]{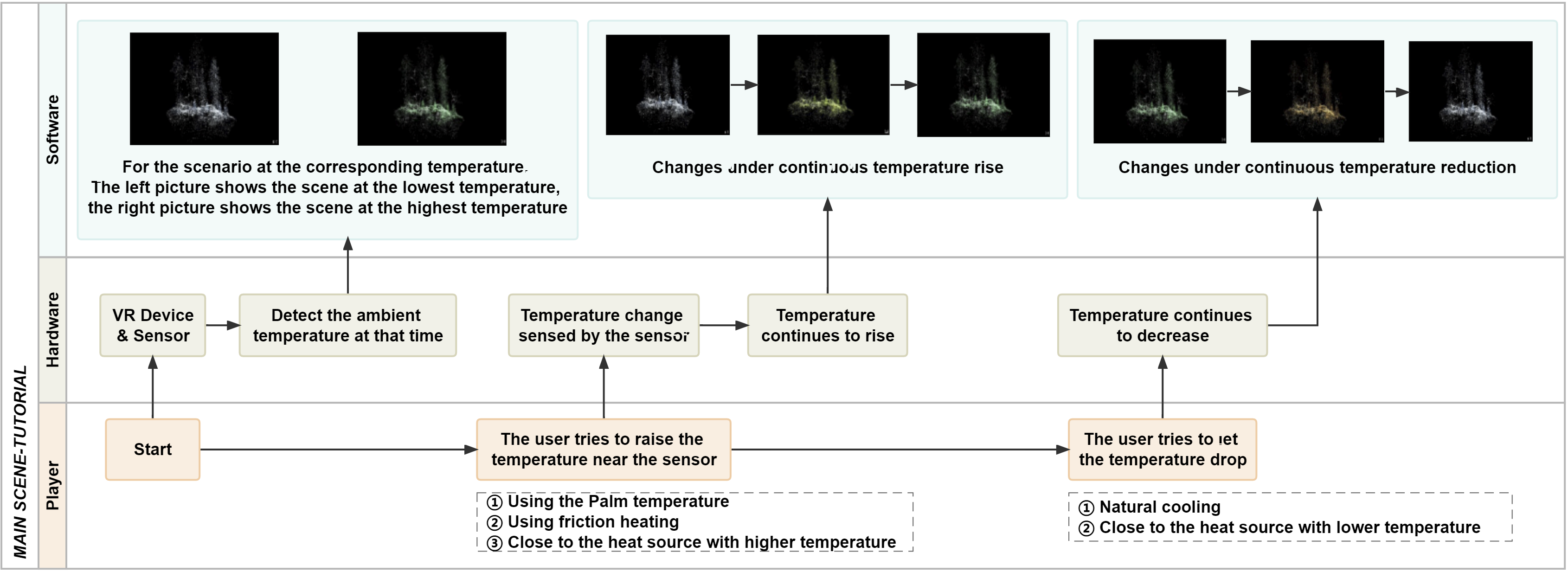}
    \caption{System flow diagram. The logic between users, hardware, and scenarios. Upon initialization, the system will display the winter scene if the ambient temperature is less than or equal to the minimum temperature (15 ℃). It will display the summer scene if the temperature exceeds or exceeds the maximum temperature (25 ℃). For temperatures between 15℃ and 25℃, the system randomly displays either the autumn or spring scene. Once initialized, the system presents a winter-spring or autumn-summer change if the temperature changes from low to high. The system presents a summer-autumn or spring-winter change if the temperature changes from high to low.}
    \Description[System flow diagram.]{The logic between users, hardware, and scenarios. Upon initialization, the system will display the winter scene if the ambient temperature is less than or equal to the minimum temperature (15 ℃). It will display the summer scene if the temperature exceeds or exceeds the maximum temperature (25 ℃). For temperatures between 15℃ and 25℃, the system randomly displays either the autumn or spring scene. Once initialized, the system presents a winter-spring or autumn-summer change if the temperature changes from low to high. The system presents a summer-autumn or spring-winter change if the temperature changes from high to low.}
    \label{fig:MainSceneTutprial}
\end{figure*}

Through exploring the original and produced virtual scenes, this research seeks to improve the user experience of VR tours in the post-pandemic era (without BIM building information models) and encourage consumers to consider how the actual and virtual worlds interact. 
Our project will be presented as an interactive VR installation, using a browser as our primary platform (utilizing AFrame for webVR). The display device will be an Oculus Quest 2 that has been modified with a variety of sensors to enable interactive tactile mechanisms for \textit{temperature}, \textit{sound}, and \textit{humidity}. 
We link the VR headset gadget with sensors and Arduino to create the final installation. The user's behavior will affect the sensor adaption parameters, which will then be transmitted to the VR headset to modify the virtual scene. For instance, if the user places a humidity sensor in a glass of water before them, the virtual scene will shift from rain to rainstorm to snow. Similarly, warming the temperature sensor with a lighter will gradually transform the virtual scene from a snowy day to spring and summer as shown in Figure~\ref{fig:teaser}.

\section{Related Work}

This section provides an overview of the related literature in three areas: 1) Virtual Reality and the Web, 2) Interactive Web and Artworks, and 3) Tangible Websites. We have a better grasp of physical networks in virtual reality thanks to the literature reviews in these three directions so that we may create interactive network art that is more immersive and enjoyable for users.

\subsection{Virtual Reality and the Web}

"Virtual Reality" \cite{schroeder1996possible} refers to a simulated or artificial environment that allows users to experience an environment outside their natural world. In recent years, VR technology has undergone rapid hardware and software advancements. For instance, Facebook's recent rebranding to Meta \cite{bbc} has pushed the concept of the metaverse to the forefront, while the global VR market's hardware sales are projected to reach 9.6 million in 2022 \cite{CNBC}. As a result, VR technology is increasingly being applied across various industries, such as education, medical treatment, and art, with more artists considering virtual space to showcase their artworks and provide immersive experiences to audiences.

The growing attention paid to VR has prompted tens of thousands of developers to invest in research, leading to the development of three main ways of VR: (1) VR software development based on the Android platform, which involves many underlying technologies, has a high threshold for getting started, and has poor cross-platform compatibility; (2) based on game engines such as Unity \cite{Unity} and UE \cite{UE}, which significantly reduces the threshold for game engine development, but still faces platform compatibility challenges; and (3) the emergence of the WebVR \cite{webVR} standard in 2016, which presents developers with new ideas. The platform compatibility attribute of the website has piqued developer interest in this development method. After several years of development, WebXR \cite{webXR} appeared in 2019, enabling developers to not only develop VR applications based on the web but also AR applications. The website hosts many excellent application cases, such as MoonRider \cite{MoonRider}, a music visualization and rhythm game developed by two people, and A-Painter \cite{A-painter}, a game that allows users to graffiti on virtual space.

\subsection{Interactive Web and Artworks}

When creating web pages, it is important to utilize user-centered design methods to enhance interactivity. Interactive content can encourage users to actively engage with the content and establish a deeper connection with the web page \cite{interactive-tool}. Traditional interactive features on websites typically include voting, evaluations, commenting, and social dialogue \cite{interactive-tool}. Interactive content promotes understanding by enabling users to actively explore information, which is more effective compared to traditional methods \cite{mcinerny2014information, spiegelhalter2011visualizing, strecher1999interactive}. It also allows users to participate and become familiar with unfamiliar subjects \cite{pidgeon2014creating}. For example, Gelman et al. \cite{gelman2013infovis} explored the role of static graphics in interactive content. Furthermore, studies have shown that adding a game mechanism to web pages can make them more appealing to users by providing an entertaining environment \cite{mayer2014research}. The development of network technology has enabled the use of two-dimensional and three-dimensional interactive content on websites, such as three-dimensional maps and games, which provide an exciting and novel experience for users. Compared to regular, static content, interactive content has numerous advantages.

However, despite the benefits of interactive content, most online artworks are still limited to static images or videos with minimal interactivity, which typically involve mouse clicks and other basic methods. This can result in a somewhat dull interaction between the artworks and the audience. With the advent of the pandemic and the rise of new-generation network technology, more artists have begun to showcase their artworks on the internet. However, the lack of interactivity may be a hindrance to a truly engaging experience for viewers.

\subsection{Tangible Websites}


The tangible web offers a new perspective on web design that brings the intangible world of the web into the physical realm, creating a stronger connection between the two. In the study conducted by Jorge Garza et al. \cite{garza2021appliancizer}, interactive elements of web pages such as buttons were transformed into tangible buttons in the real world. Similarly, Ashrith Shetty et al. \cite{shetty2020tangible} created a touchable web page that allowed visually impaired individuals to experience the layout of web pages with their hands. Jiasheng Li et al. \cite{li2022tangiblegrid} proposed a device that provided real-time tangible feedback to allow visually impaired users to understand the layout of web pages.

The tangible web offers a unique way to interact with web pages in two ways. Firstly, it materializes the intangible web, allowing users to physically touch it. Secondly, physical interactions in the real world can affect the content of the web page. This opens up new possibilities for interaction and creativity in presenting artworks on the web. By utilizing a physical system, changes in the physical world can affect the content on the web page, providing more dynamic and engaging interactions. This approach can be applied to the presentation of artworks on the web, making them more interactive and visually appealing to users.

\section{System Design and Implementation}

The proposed artwork leverages temperature sensing and webVR to offer the user a unique experience where the physical world directly affects the web environment. This section details the general architecture behind 

\subsection{System Design}

\begin{figure}[t]
    \centering
    \includegraphics[width=\linewidth]{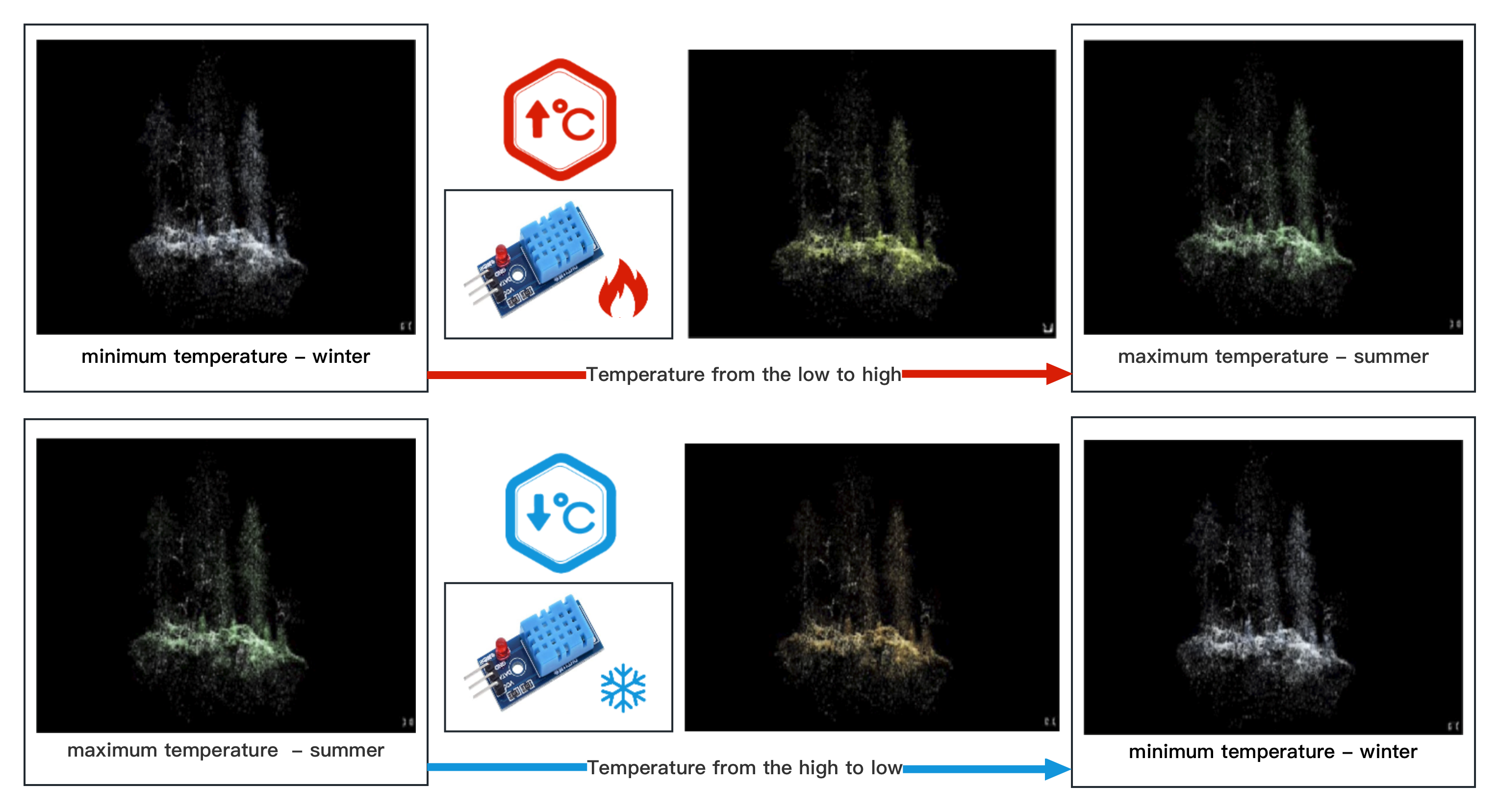}
    \caption{Scene Design. The transition of the scene from high temperature to low temperature and from low temperature to high temperature.}
    \Description[Scene Design.]{The transition of the scene from high temperature to low temperature and from low temperature to high temperature.}
    \label{fig:SceneDesign}
\end{figure}

The proposed artwork aims to create an immersive experience that responds to environmental changes detected by sensors (see Figure~\ref{fig:MainSceneTutprial}). Users are required to wear VR devices equipped with sensors to participate in the experience. These sensors record the temperature over a set period and calculate the average values. Based on the temperature recorded, the corresponding scene is displayed (see Figure~\ref{fig:SceneDesign}).
As a proof-of-concept system, we have implemented two scenes corresponding to the lowest and highest temperature, respectively. Two temperature thresholds have been set, triggering the corresponding scene's display. The system displays the low-temperature scene below the lower threshold and the high-temperature scene above the higher threshold. The system selects a random scene if the temperature is between both thresholds. 
Users can manipulate the temperature around the sensor in several ways, including increasing it with their palm temperature, generating friction heat, or moving the sensor closer to a heat source (see Figure~\ref{fig:SceneDesign}). This causes the scene to transition from low to high temperature. Alternatively, users can decrease the temperature through natural cooling or by placing the sensor close to a cold source. This causes the scene to transition from high to low temperature.


\subsection{Implementation}

\begin{figure}[t]
\centering
\includegraphics[width=\linewidth]{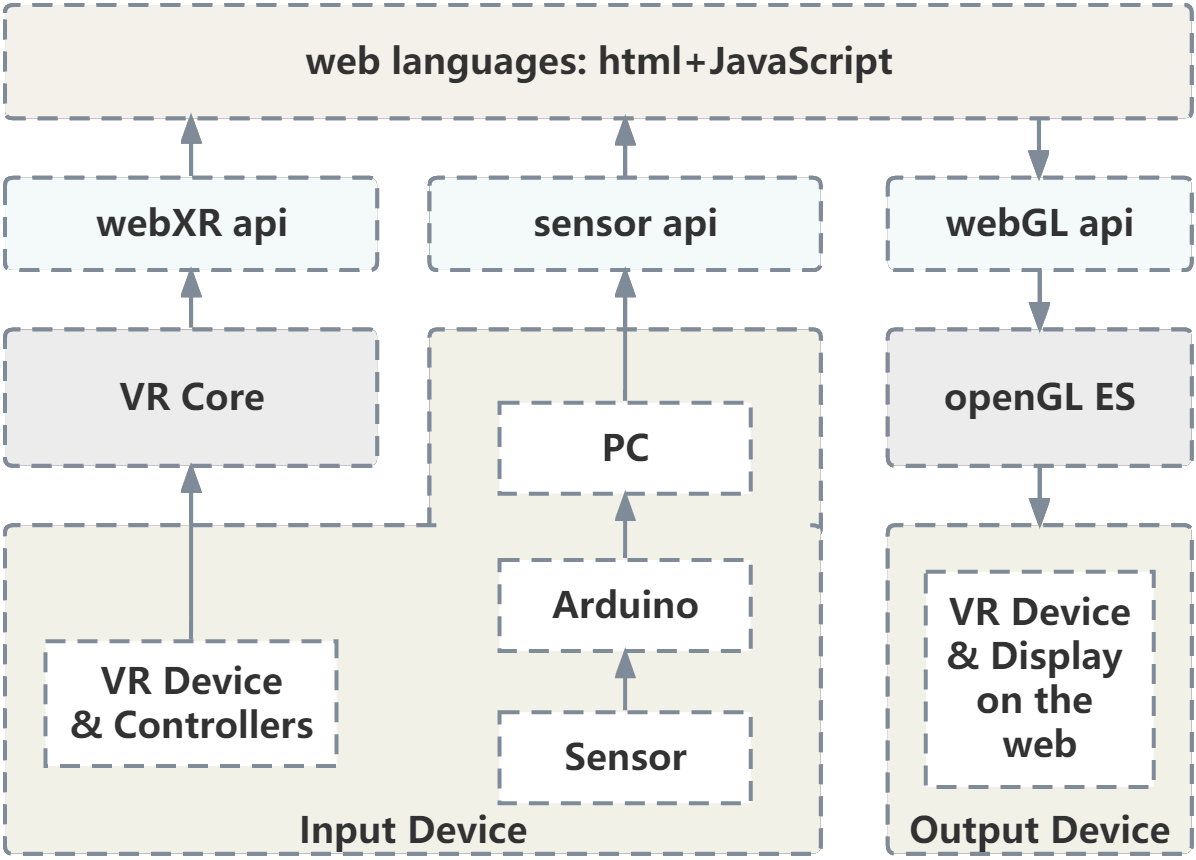} 
\caption{Technical architecture. This figure describes how the input signal is compiled into web language and then displayed in the web page of VR device. We convert VR and sensor applications into web language through webXR api and sensor api and then present 3D scenes on the web through webGL api to let users experience the web in VR devices.}
\Description[Technical architecture.]{This figure describes how the input signal is compiled into web language and then displayed in the web page of VR device. We convert VR and sensor applications into web language through webXR api and sensor api and then present 3D scenes on the web through webGL api to let users experience the web in VR devices.}
\label{tech}
\end{figure}

The system is implemented as a WebVR experience that collects data from a temperature probe located near the user. We utilize an Arduino microcontroller to receive and process the data from the probe, which is then sent to the PC via the serial port. The PC processes the data to control the display of the VR device (as shown in Figure \ref{tech}). We have experimented with two sensors: a humidity sensor (DHT11) and a flame sensor (infrared receiver module). The DHT11 is a digital temperature and humidity sensor with a measurement range of 0 °C to 50 °C, suitable for most scenes. We connect the DTH11 output to the digital IO port of the Arduino. Additionally, the flame sensor (infrared receiver module) detects flames through its unique infrared receiving tube and converts the flame's brightness into a signal output by digital or analog value. While Although it doesn't detect changes in temperature, this feature opens up new possibilities for our system's interactions. However, we caution against using fire in hazardous environments.

The Arduino communicates with the computer via a serial port. We write code to receive serial port data continuously, with a baud rate of 9600 and 8 data bits. After data processing, the VR device's display is controlled through wired or wireless links. Once the system is connected, users can experience changes in the virtual world through sensors.

\begin{figure}[t]
    \centering
    \includegraphics[width=\linewidth]{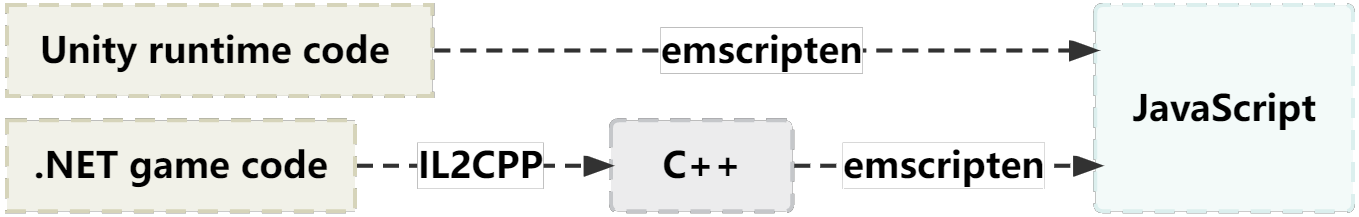}
    \caption{Workflow in Unity. This figure depicts how unity cross-compiles the runtime code and .NFT game code into web language.}
    \Description[Workflow in Unity.]{This figure depicts how unity cross-compiles the runtime code and .NFT game code into web language.}
    \label{fig:workflow}
\end{figure}

We implemented our system using Unity, which compiles runtime code into JavaScript files using the emscripten compiler toolchain. The .NET game code is compiled into C++ files through IL2CPP and then into JavaScript files using the emscripten compiler toolchain (as shown in Figure \ref{fig:workflow}).

As the temperature sensor transmits different values to the computer, the scene users see in the VR device will also change (see Figure \ref{fig:MainSceneTutprial} \& Figure \ref{fig:teaser}). When the temperature drops to the lowest, users will see the snowy forest scene, and as the temperature rises, the scene users see will change from white to green. This symbolizes the melting process of ice and snow. When the temperature is the highest, users will see the emerald forest. However, as the temperature decreases, the emerald forest will gradually turn yellow and white, symbolizing the transition from midsummer to winter.

\begin{figure}[t]
\centering
\includegraphics[width=\linewidth]{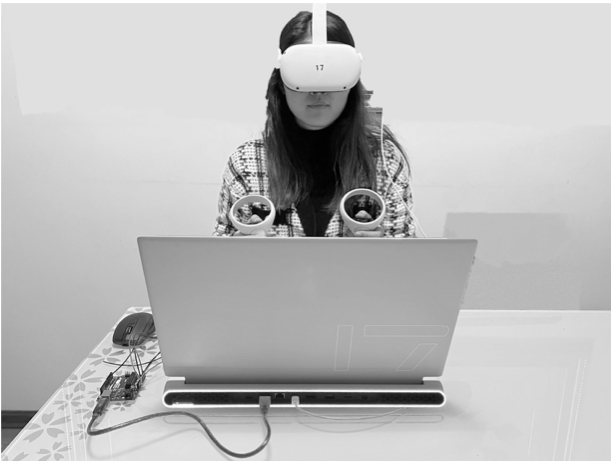} 
\caption{A user interacting with our system. The user can control the controller and sensor at the same time and see the scene changes.}
\Description[A user interacting with our system.]{The user can control the controller and sensor at the same time and see the scene changes.}
\label{image1}
\end{figure}


\section{User Feedback}
We presented our work to various audiences, including artists and designers (identified as A1, A2, and A3), curators (identified as C1, C2, and C3), and general audiences (identified as G1, G2, G3, G4, G5, and G6). These engagements provided valuable feedback and insights from individuals with diverse perspectives and expertise. Also, we have obtained approval from the  Ethics Committees of the respective organizations for studies.
\subsection{Audience Responds Positively to Innovative Virtual Environment}
Most of the audience found our virtual environment innovative and aesthetically pleasing. However, a few viewers, specifically C1 and A3, initially struggled to navigate the virtual environment and experienced a feeling of being lost. After we explained our design process and provided examples, they better appreciated the environment's aesthetic qualities. Several viewers, including C1, C3, G2, G3, and A1, were particularly impressed with the scene changes in the virtual environment, citing the bright colors and magical transformation process as highlights. In particular, G1 and C1 found the winter environment visually comforting and relaxing. C1 even suggested the potential for designing souvenirs based on the visual elements of the scene. Finally, C5 expressed appreciation for the interactive elements of the environment that triggered scene changes.

\subsection{Interactive Art Installations}

The audience appreciated the virtual scenes generated by the point clouds, which helped them better understand the new story behind the scenes. For instance, G4 commented that the temperature change that caused the seasons to shift in the scene made sense. It mirrored real-life behavior in a virtual environment, making exploring it enjoyable. Similarly, G2 appreciated the interactive aspect of the virtual scenes, which he had not fully realized before. This interactive approach piqued his interest and made him eager to explore further.

Artists A1 and A2 praised this new interactive form for enabling them to create new artworks. They noted that they could use this system to develop interactive art installations, allowing visitors to engage deeply with their works and experience a sense of connection. This system will undoubtedly prove helpful in their future artistic endeavors.
\subsection{Creating Art that Responds to Personal Behavior}



Interactive art has become the mainstream form of art today, and exploring the interaction between reality and virtual is highly significant for artists and designers. The experience of using this system provides them with new ideas in the creative process. However, there is a need to increase gesture interaction by incorporating a motion capture system.

Curators are more interested in novel art forms that can attract players, but they also prioritize the convenience of equipment management. Therefore, they require an integrated and easy-to-manage system.

For general audiences, the opportunity to influence changes in artwork through personal behavior is a new and exciting experience, allowing them to feel empowered and in control like a deity. However, if the sensor is too simple or limited, they may gradually lose interest and require ongoing stimulation to maintain engagement. Therefore, it is highly recommended to incorporate more random elements to enhance the experience. 
\section{Discussion}
This paper presented a creative approach to improving immersive web experiences by integrating physical world sensors, like temperature and humidity, with Web XR applications. The interactive virtual reality system "Tangible Web" helped us transfer real-world data to a virtual reality world, empowering web VR with additional capabilities and enabling users to interact with the VR environment more deeply and tangibly. 


Through user feedback, we have gained valuable insights into how our system could cater to different interested parties. Our system has been well-received by artists, who appreciate its potential to provide new creative opportunities. This is consistent with past research highlighting the importance of incorporating new technologies in creative processes \cite{jeon2021fashionq, rees2015building}. Curators have also given positive feedback, noting the system's novelty and highlighting the importance of system integration, ease of storage, and management. This aligns with previous research, emphasizing the importance of user-friendly technology in museum and exhibition settings \cite{harms2001evaluating, dumitrescu2014creating}. Finally, the audience responded positively to the system's playability and their ability to participate in the artwork. However, some viewers have suggested that they would like the system to be more open, allowing them to interact with preset works and participate in creating new artwork. Thus, our system has the potential to be further enhanced to meet the needs and expectations of a broader range of stakeholders while also opening up new opportunities for artists to explore and innovate in their respective fields.

Our research participants emphasized the importance of incorporating the element of surprise in interactive art installations. This idea aligns with the findings of a study on the effect of uncertainty on positive events \cite{wilson2005pleasures}, which found that uncertainty following a positive event can prolong the pleasure it causes. By including random elements that create unexpected and novel connections between the digital and physical worlds, the overall user experience could be significantly enhanced. 

\subsection{Different Application Domains}
The potential applications of the Tangible Web extend far beyond museums and exhibitions, encompassing diverse fields such as gaming, education, and healthcare.

For example, in a game context, incorporating physical sensors could add realistic elements like wind and temperature, enhancing the gameplay experience and creating more epic and emotional scenes for gamers. In adventure games, the motion capture system proposed by some artists and designers could be used to create more realistic and responsive character movements, adding to the game's immersion. Moreover, designers and artists can employ well-known personality traits theories such as the Big Five and models like Brain Hex to tailor interactive art installations to the expectations of their audiences, taking into account their personalities and preferences \cite{gameplay23}. By doing so, they can create more immersive content that resonates with them deeper, making it more memorable and impactful.

In an educational context, Tangible Web provides an interactive and engaging way to learn by responding to physical sensors. For instance, a science lesson on plant growth could use sensors to monitor temperature, humidity, and light levels, affecting the illustrations displayed in a virtual garden. Students can interact with the garden, observing how environmental changes impact plant growth and learning about the science behind it. This approach encourages active learning and helps students to develop a deeper understanding of scientific concepts and critical thinking skills that can benefit their future studies and careers \cite{hadi2021student}. 


In a healthcare context, Tangible Web could be utilized to create interactive therapy tools that respond to physical sensors. For example, a virtual reality experience that guides patients through breathing exercises could incorporate sensors that measure heart rate and respiration. The virtual environment would adjust in real-time to provide feedback and help patients achieve optimal breathing patterns. This could be especially useful for anxiety, stress, or chronic pain patients.

\subsection{Self-Determination Theory and Tangible Web}
Self-determination theory (SDT) is a motivational theory that can be applied to the design of interactive art installations. The theory proposes that humans have three basic psychological needs: \textit{autonomy}, \textit{competence}, and \textit{relatedness} \cite{tyack2020self, gameplay23}. Autonomy refers to the need to feel in control of one's actions and decisions. Competence refers to feeling capable and effective in one's actions. Relatedness refers to the need to feel connected to others and have a sense of sharing and belonging.

Incorporating Self-Determination Theory (SDT) principles into interactive art installations could enhance users' motivation, and engagement \cite{hadi2022users}. For instance, granting users a sense of autonomy by allowing them to control the artwork's behavior through their behavior can increase their motivation to interact with it. Designers can also increase users' motivation and engagement by providing opportunities to demonstrate competence through successful interactions with the artwork and creating new connections between physical and virtual reality. Moreover, designers can foster a sense of relatedness by encouraging collaborative interactions or shared experiences, empowered by shared feelings in both virtual and real worlds. Therefore, by combining SDT principles and new features enabled by Tangible Web, designers can provide users with a more satisfying and meaningful experience when engaging with interactive art installations.

\section{Conclusion and Future Work}

The question of enhancing a tour's user experience and enjoyment using digital technology and interaction frequently arises in the modern era, mainly when epidemics restrict our ability to travel. When we take virtual scenario tours, we tend to concentrate on how similar the virtual and real-life scenarios are rather than how the virtual scenario enhances the real one and vice versa. We may see the splendor of the scene in front of us and the beautiful sensation of the changing seasons and weather in the virtual scene. However, to fully engage us, interactive technology must be incorporated into the design of the virtual experience. This strategy has been shown to impress audiences, making it a promising approach to enhancing the user experience and enjoyment of virtual tours \cite{walczak2006dynamic}.

Current VR technology research and development focus on accurately transferring the five sensations of reality, including haptics, into the virtual environment. Regardless of the availability of technology, the question remains: how can we minimize the difference between the real and virtual worlds? To achieve this goal, it is necessary to reflect on the potential of various technologies to create an immersive experience that is contextually relevant to the audience while also paying attention to the professional and ethical aspects of the project \cite{germak2021augmented}. Visitors are fascinated when stories are told that they can relate to and when they feel like the protagonists of a new reality \cite{han2018user}.


The scalability of VR technology allows for continuous content upgrades and the development of interactive experiences, providing opportunities to enhance the tourist experience and enjoyment in the post-pandemic period. Interactive technologies like the "Tangible Web" can create immersive environments that bring virtual and physical realities closer together, blurring the lines between the two. These environments can be customized for various applications, offering a range of experiences that meet the needs of diverse audiences. By leveraging these advancements, we can create more engaging and captivating virtual tours that offer unique and unforgettable experiences for users.


Future work will focus on developing applications for numerous sensors to adapt them to a more robust interactive experience. It is also a priority to develop modular creator tools that make it possible to customize the addition of corresponding sensor modules based on the user's needs. Additionally, an interactive asynchronous communication system will be developed to boost user viscosity in the online community. The aim is to translate this experience to the metaverse to enhance the interactive experience and strengthen the link with reality. Finally, eco-art in the metaverse is another sustainable direction to explore, such as intelligent communities, smart cities, and the sustainability of virtual creative production.


\section{Acknowledgement}

This research was partially supported by the MetaUST project from HKUST(GZ) and the  FIT project (Grant No. 325570) from the Academy of Finland.
\bibliographystyle{ACM-Reference-Format}
\bibliography{sample}

\appendix

\end{document}